\documentclass[aps,pra,reprint,superscriptaddress,longbibliography]{revtex4-2}
\usepackage{graphicx} 
\usepackage{amsmath}
\usepackage{xcolor}
\usepackage{braket}
\usepackage[colorlinks=true,citecolor=blue,linkcolor=blue,urlcolor=blue]{hyperref}
\usepackage[capitalise]{cleveref}
\usepackage{natbib}

\newcommand{\e}{\epsilon}

\begin{document}

\title{Quantum solitons and their quantum walks in transmon arrays}

\author{Ben Blain}
\affiliation{Quantum Research Center, Technology Innovation Institute, Abu Dhabi 9639, UAE}
\affiliation{School of Engineering, Mathematics and Physics, University of Kent, Canterbury CT2 7NH, United Kingdom}

\author{Giampiero Marchegiani}
\email{giampiero.marchegiani@tii.ae}
\affiliation{Quantum Research Center, Technology Innovation Institute, Abu Dhabi 9639, UAE}

\author{Luigi~Amico}
\affiliation{Quantum Research Center, Technology Innovation Institute, Abu Dhabi 9639, UAE}
\affiliation{Dipartimento di Fisica e Astronomia, Via S. Sofia 64, 95123 Catania, Italy}
\affiliation{INFN-Sezione di Catania, Via S. Sofia 64, 95127 Catania, Italy}

\author{Gianluigi Catelani}
\affiliation{Institute for Theoretical Nanoelectronics (PGI-2),Forschungszentrum J\"ulich, 52428 J\"ulich, Germany}
\affiliation{Quantum Research Center, Technology Innovation Institute, Abu Dhabi 9639, UAE}

\begin{abstract}
Superconducting qubits are artificial atoms whose spectra and interactions can be engineered through appropriate circuit design, a versatility that can be exploited for quantum simulation. We theoretically investigate a linear array of capacitively coupled transmons, effectively described by a Bose-Hubbard Hamiltonian with attractive interaction. We revisit the discrete-soliton nature of the lowest-energy band of the spectrum, and identify spatially localized quantum solitons. The solitonic character of these states is revealed through their time evolution, which displays a quantum interference pattern, or quantum walk, highlighting their composite nature. We discuss protocols for preparing spatially localized quantum solitons that are compatible with current state-of-the-art tunable-transmon circuits. Our results demonstrate that superconducting circuits provide a promising and experimentally accessible platform for the investigation of quantum soliton physics.
\end{abstract}

\date{\today}

\maketitle

\section{Introduction}

Quantum simulation has emerged as a cornerstone of modern many-body physics, thanks to controllable platforms enabling the exploration of regimes that remain computationally challenging for classical algorithms~\cite{buluta_quantum_2009,Georgescu2014QuantumSimulation,Altman_2021,Fauseweh_2024}. Among the most promising architectures, superconducting circuits -- and specifically arrays of transmon qubits -- have demonstrated the versatility required to emulate complex lattice models~\cite{houck2012chip}. While the higher energy levels of weakly anharmonic transmons are traditionally viewed as a source of ``leakage'' that hinders high-fidelity quantum gates~\cite{Blais:2020wjs}, they can also be harnessed as a valuable resource. By using these higher levels,  transmons can be treated as qutrits or qudits, expanding the local Hilbert space for more complex simulations~\cite{Goss2022}.
A particularly compelling application of superconducting qubit arrays is the realization of the Bose-Hubbard lattice model with attractive interactions~\cite{hacohen2015cooling,fedorov2021photon,mansikkamaki_phases_2021,mansikkamaki_beyond_2022,berke_transmon_2022,Blain:2022qur,Chirolli_2025}. In contrast to the many-body systems with repulsive particle-particle interaction, where the system tends to minimize local occupation to avoid energy penalties, the low-energy quantum dynamics of attractive bosons is characterized by states in which particles  ``pile up'' at specific lattice sites. Such a feature leads to large local Hilbert spaces which, together with the strong correlations, pose a computational overhead that is difficult to overcome with classical simulation methods~\cite{aaronson_computational_2013,gard_inefficiency_2014}. Therefore, quantum simulation is particularly relevant for correlated bosonic particles with attractive interactions.

The ``piling up'' of particles is not evident in simple observables such as the local density of low-lying eigenstates; however, it manifests itself, for instance, in the decay of the density-density correlation of the ground state~\cite{naldesi_rise_2019}. This feature indicates that the ground state can be thought of as a superposition of solitons. In Ref.~\cite{naldesi_rise_2019}, it was  proposed to introduce a pinning site to select a soliton out of the ground state, and the short-time evolution after releasing the pinning was studied. More recently, we remarked that this initial evolution displays some properties of quantum walks~\cite{Blain:2022qur}. In this work, we revisit quantum solitons in the attractive Bose-Hubbard model. We discuss how  spatially localized quantum solitons can be defined, and propose a protocol suitable for current superconducting hardware for preparing a localized soliton with high fidelity. We then study the quantum walk of a soliton and quantitatively compare it with that of a single particle and of a boson stack.

Our interest in quantum walks is twofold. From a fundamental perspective, quantum walk interference patterns are a vivid manifestation of particle-wave duality. For indistinguishable multi-particle walks, they give evidence for correlations due to interactions~\cite{Lahini2012,Preiss2015,CaiPRL127,oliveira_cooperative_2024}. In this light, quantum solitons in attractive Bose–Hubbard arrays can be viewed as composite walkers made of bound particles whose correlated motion enhances the resulting interference pattern. From the point of view of potential applications, it is established that quantum
walks provide a natural framework for quantum search algorithms, where interference on a graph can yield quadratic speedups over classical random-walk searches~\cite{childs_spatial_2004}; they can even realize universal quantum computation~\cite{Childs_Science2013,qiang_quantum_2024}. More importantly in the context of quantum solitons, multi-particle walks could enable quantum-enhanced measurement~\cite{Compagno2017NOON,CaiPRL127}. 

This article is organized as follows. In Sec.~\ref{sec:quantumSolBH} we construct and characterize spatially localized quantum solitons in the Bose-Hubbard model. In \cref{sec:TransmonSoliton} we discuss protocols and present theoretical results for the preparation of a quantum soliton in 1D transmon arrays. In \cref{sec:single-particle} we perform quantitative analyses on the differences between a quantum soliton and a boson stack, including the quantum soliton prepared with our process in \cref{sec:TransmonSoliton}. Finally, in \cref{sec:conclusions} we summarize our analysis and results.

\section{Quantum solitons in the Bose-Hubbard model}
\label{sec:quantumSolBH}

We consider a system of interacting bosons in a one-dimensional array of $M$ sites, modeled by a Bose-Hubbard (BH) Hamiltonian
\begin{align}
    \frac{\hat{\mathcal{H}}_{\rm BH}}{\hbar} = &\sum_{i=1}^{M-1} J_{i,i+1} \left(\hat{b}_i^\dagger \hat{b}_{i+1} + \hat{b}_{i+1}^\dagger \hat{b}_{i}\right) + \frac{U}{2} \sum_{i=1}^M \hat{n}_i (\hat{n}_i - 1) \nonumber\\
    &+ \sum_{i=1}^M\mu_i(t) 
    \hat{n}_i\ ,
    \label{eq:BHm}
\end{align}
where $U$ is the interaction strength, which we assume attractive ($U<0$), $\mu_i$ ($i=1,\dots, M$) are the site-dependent chemical potentials, and $J_{i,i+1}$ are the neighboring-site hopping coefficients. In Eq.~\eqref{eq:BHm},  $\hat{b}_i^\dagger$ ($\hat{b}_i$) denotes the creation (annihilation) operator acting on site $i$, obeying the commutation relation $[\hat{b}_i, \hat{b}_j^\dagger] = \delta_{ij}$, and $\hat{n}_i = \hat{b}_i^\dagger \hat{b}_i$ counts the number of bosons on the site. Since the Hamiltonian commutes with the number operator $\hat{N}=\sum_{i}\hat{n}_i$, we can treat each subspace with fixed number of bosons $N$ separately. For the state preparation in a transmon array addressed below in Sec.~\ref{sec:TransmonSoliton}, we assume that the chemical potentials can be dynamically tuned in time $t$.
\begin{figure}
    \centering
    \includegraphics[width=\linewidth]{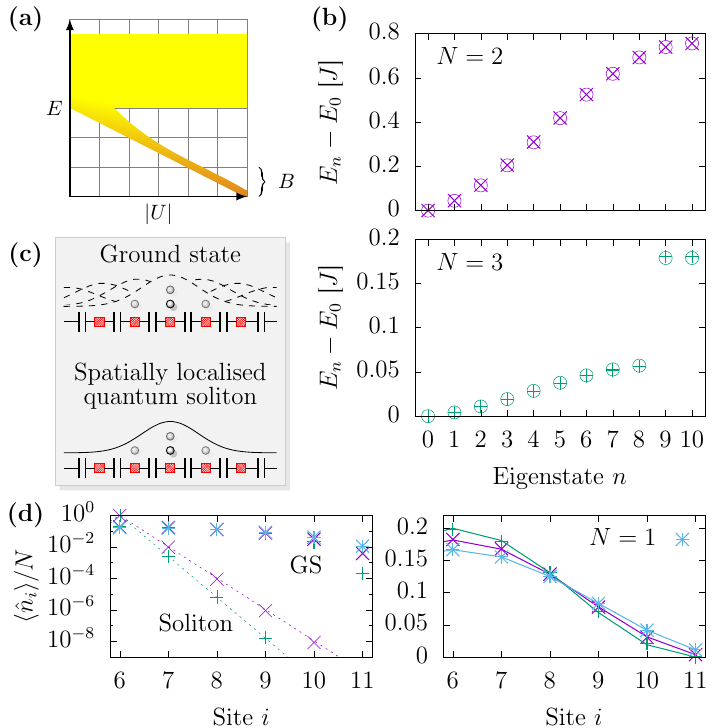}
    \caption{Quantum solitons in the BH model. \textbf{(a)} Schematic showing the spectrum of Eq.~\eqref{eq:BHm} system as a function of the interaction strength for homogeneous hopping and chemical potential. As the interaction increases, a  clear energy gap develops, with a soliton band (here labeled as $B$) composed of $M$ states detaching from the rest of the spectrum.
    \textbf{(b)} Sorted eigenvalues of the Bose-Hubbard model in the soliton band. The dashed line shows a numerical approximation for $N=2$ -- see text for details.
    \textbf{(c)} Schematics showing the ground state of the BH model (top) and a spatially localised quantum soliton (bottom).
    \textbf{(d)} Per-site density of the ground state (GS, solid connecting lines) of the Bose-Hubbard model and a spatially localised quantum soliton (dotted connecting lines) for $N=2,3$ total particles, both plotted in logscale (left) and GS data in linear scale (right). The solid connecting lines show the expressions as written in the text for integer $i$, while the dashed are given by $\langle \hat n_i\rangle/N = \exp(-|i-6|/\xi_N)$ with $\xi_N$ as given in \cref{eq:xiN-theory}. Parameters in panels (b) and  (d) are $U=-10J$, $M=11$.
    }
    \label{fig:BHm-eigenenergies-density}
\end{figure}
In a uniform system, with site-independent chemical potential $\mu_i=\mu$ and hopping coefficients $J_{i,i+1}=J$, the eigenstates and their spectrum are determined by the interplay between the hopping and the interaction strength.
At zero interaction, $U=0$, the BH model reduces to a tight-binding model that can be diagonalized by Fourier transformation. The resulting eigenstates can be constructed from single-particle states with well-defined wave number $k_n=\pi (n+1)/(M+1)\ (n=0\ldots M-1)$, and single-particle energy $E_n=2J\cos(k_n)$. As $|U|$ increases, the $M$ states in which all the bosons predominantly occupy the same site become favorable in energy due to the attractive interaction. When the size of the interaction strength exceeds a critical value $U_C$ which depends on the number of particles, $U_C\simeq J[2N/(N-1)!]^{1/(N-1)} $~\cite{naldesi_rise_2019,Blain:2022qur}, these states become distinctly separated from the remaining eigenstates by an energy gap~\cite{boschi_bound_2014,valiente_two-particle_2008}. Wave-packets constructed from these states have been shown to be stationary solutions of the discrete nonlinear Schr\"odinger equation in the correspondence limit $N\gg 1$, and so the band formed by these states is called the ``soliton band''~\cite{Scott:1994bm}. The evolution of the spectrum with the interaction strength and the soliton band are schematically depicted in Fig.~\ref{fig:BHm-eigenenergies-density}(a). 

In the opposite limit characterized by a small number of sites and few bosons, it is natural to seek a quantum version of the classical solitons. This setup is of interest, for instance, for quantum simulation with superconducting transmon qubits where bosons correspond to plasmon excitations localized in the transmons. Typical transmons are in the regime of strongly attractive bosons $-U\gg J$, but the regime of intermediate interaction strength $|U| \gtrsim U_C$ can be engineered.
In the strongly attractive regime, the soliton band can be effectively described in terms of a Hamiltonian for composite particles with $N$ bosons on the same site: these composite particles can hop with an effective hopping coefficient 
\begin{equation}\label{eq:Jtilde_def}
\tilde{J}\simeq J \frac{N}{(N-1)!}\left(\frac{J}{|U|}\right)^{N-1} 
\end{equation}
and a chemical potential profile -- symmetric with respect to the array center -- in which consecutive sites have a mismatch $\tilde{\mu}_i-\tilde{\mu}_{i+1}\simeq NJ[J/U(N-1)]^{2i-1}$ due to boundary effects (see Ref.~\cite{mansikkamaki_beyond_2022} and Appendix~\ref{app:variational}). Therefore, the energy spectrum of the soliton band depends on both $N$ and $M$.

In Fig.~\ref{fig:BHm-eigenenergies-density}(b), we display the energy dispersion in the soliton band for $N=2$ (top panel) and $N=3$ (bottom panel), obtained by numerically computing (with exact diagonalization) the spectrum of Eq.~\eqref{eq:BHm} for uniform chemical potential $\mu_i=\mu$ and hopping $J_{i,i+1}=J$ for $U=-10 J$ and $M=11$. For $N=2$, the energies of the soliton band can be estimated with good accuracy by adapting the known exact expression for periodic systems~\cite{Scott:1994bm} to the chain case, $E_n \simeq \text{sgn}(U)\sqrt{U^2+16 J^2 \cos^2(\tilde{k}_n/2)}$ for $n=0\ldots M-1$ (shown by the dashed line). The wave-numbers $\tilde{k}_n=\pi (n+1)/M$ account for the open boundaries and the presence of an effective chemical potential at the boundary $\tilde\mu_{1}=\tilde\mu_{M}\simeq \tilde J$~\cite{Banchi2013Spectral}. For $N=3$, the chemical potential mismatch at the edges is much larger than the effective hopping. Thus, we can approximate the lowest $M-2$ levels in the soliton band with the dispersion of an $N$-boson particle with hopping $\tilde{J}$, i.e.,
$E_n\simeq - 2\tilde{J}\cos(k_n)$ with $k_n=\pi (n+1)/(M-1)$ ($n=0\ldots M-3)$, as if the chain effectively has only $M-2$ sites. The upper two levels in $B$ are approximately separated by an energy $\tilde{\mu}_1=3J^2/2|U|$  from the center of the band of the lowest $M-2$ levels, i.e. $E_{m-2}\simeq E_{m-1}\simeq E_{(M-3)/2}+3J^2/2|U|$.
For larger $N$ (not shown) the number of sites in the soliton band which are separated in energy due to boundary effects amounts to $2(\lceil N/2\rceil-1)$~\cite{mansikkamaki_beyond_2022}.

We next discuss some properties of the ground state of the BH Hamiltonian.
\Cref{fig:BHm-eigenenergies-density}(d) shows the expectation value of the (normalized) local number operator $\hat{n}_i/N$ in the ground state of the BH model  for $N=2$ and $N=3$ bosons plotted on a logarithmic scale (left) and linear scale (right). While the occupation of the edge sites is suppressed when increasing the total bosonic number due to the edge effects discussed above, the overall density distribution depends only weakly on $N$ and $U$. In fact, the density profile is mostly related to the finite size of the chain, as confirmed by the density distribution for a single particle ($N=1$, or equivalently $U=0$),  $\braket{\hat{n}_i}|_{U=0}/N=2\sin[\pi i/(M+1)]^2/(M+1)$ (plotted in cyan solid lines). For $N=2$, we can approximate $\braket{\hat{n}_i}/N\simeq 2\sin[\pi (i-1/2)/M]^2/M$ (purple solid lines); this expression accounts for the modified wave-numbers $\tilde{k}_n$ due to the edge chemical potentials, and it also displays (thanks to the shift $i\to i-1/2$) the appropriate symmetry. For $N=3$, we can neglect in first approximation the edge sites' occupation and write the density in the remaining sites as if the chain consists effectively of $M-2$ sites, $\braket{\hat{n}_i}/N\simeq 2\sin[\pi (i-1)/(M-1)]^2/(M-1)$ (green solid lines). The three density profiles are similar and do not give a clear signature of the nature of the ground state; in contrast, for $N>1$ the density-density correlations $\braket{\hat{n}_i\hat{n}_j}$ were shown to be exponentially decaying in Ref.~\cite{naldesi_rise_2019}. More generally, each eigenstate in the soliton band can be expressed as a superposition of spatially localized quantum solitons~\cite{naldesi_rise_2019,Blain:2022qur,Scott:1994bm,Lai1989QuantumTheorySoliton}, as in the schematics of Fig.~\ref{fig:BHm-eigenenergies-density}(c). 
\subsection{Spatially localized quantum solitons}
\label{sec:spatiallyLocalized}
For spatially localized states, 
the expectation value of the occupation number is expected to decay exponentially with the distance from the localization site $i$, $\braket{\hat{n}_j}\simeq\braket{\hat{n}_i}e^{-|i-j|/\xi_N}$ for $|i-j| \gg \xi_N$, where $\xi_N$ is the decay length. Similar exponential decay occurs for the number-number correlation $\braket{\hat{n}_i\hat{n}_j}\simeq\braket{\hat{n}_i^2}e^{-|i-j|/\xi_C}$, with correlation length $\xi_C\simeq\xi_N$.  While spatially localized states can be heuristically created  as the ground state of the BH model in the presence of  a  negative chemical potential detuning (pinning) at a site~\cite{naldesi_rise_2019}, the analysis shows that such states  generally have a nonzero overlap with eigenstates of the uniform (unpinned) system laying outside of the soliton band~\cite{Blain:2022qur} and a pinning-dependent decay length (see \cref{app:variational}). With the goal of eliminating the overlap, below  we suggest three different protocols to construct spatially localized solitons around site $i$ by using soliton band states only.

\paragraph{Projected boson stack:}
Consider the Fock state with $N$ particles in site $i$, $|N_i\rangle=(\hat b_i^\dagger)^N|0\rangle$, where $|0\rangle$ is the vacuum state, which is by construction the narrowest possible state; following Ref.~\cite{mansikkamaki_beyond_2022}, we refer to such a state as a boson stack. This state is in general a superposition of eigenstates within and outside the soliton band (see Ref.~\cite{Blain:2022qur} and the next section). Thus,
we consider the projection of this state onto the soliton band of the BH model and normalize the resulting state: 
\begin{equation}
 |\psi_{P}^{(i)}\rangle=\frac{\mathcal{\hat P}_B |N_i\rangle}{\langle N_i|\mathcal{\hat P}_B |N_i\rangle} \,,
 \label{eq:projStack}
\end{equation}
where $\mathcal{\hat P}_B$ is the projector onto the lowest $M$ eigenstates of the system (we recall $\hat{P}_B^2=\hat{P}_B$). We refer to the states $|\psi_{P}^{(i)}\rangle$ as projected boson stacks. 

\paragraph{Single-stack soliton:}
In the second protocol, we construct a linear superposition of the eigenstates $\ket{\phi_n}$ ($n=0,...M-1$) in the soliton band
\begin{equation}
|\psi_S^{(i)}\rangle = \frac{\sum_{n=0}^{M-1} c_n^{(i)} |\phi_n\rangle}{\sum_{n=0}^{M-1} |c_n^{(i)}|^2}\,,
\label{eq:singleStackSoliton}
\end{equation}
which has a non-zero overlap only with a selected boson stack on site $i$. The coefficients $c^{(i)}_n$  can be found solving the following linear system: 
\begin{equation}
\sum_{n=0}^{M-1}\langle N_l | \phi_n\rangle c_n^{(i)} =\delta_{l,i},\quad l=1,...,M.
\end{equation}
We name the state $|\psi_S^{(i)}\rangle$ a single-stack soliton.

\paragraph{Narrowest soliton:}
In the third protocol, we consider the superposition of the states in the soliton band 
\begin{equation}
|\psi_O^{(i)}\rangle = \sum_{n=1}^M \tilde{c}^{(i)}_n |\phi_n\rangle\,,
\label{eq:narrowestSoliton}
\end{equation}
in which  the coefficients $\tilde{c}^{(i)}_n$ are found minimizing the width~\cite{naldesi_rise_2019,Blain:2022qur,boschi_bound_2014} of the quantum state 
\begin{equation}\label{eq:R2}
R[|\psi\rangle] = 
\sqrt{\frac{1}{N}\sum_{l=1}^M \langle \psi|\hat n_l|\psi\rangle (l-i)^2}
\,,
\end{equation}
under the normalization constraint $\sum_{n=1}^M|\tilde{c}^{(i)}_n|^2 = 1$. For the optimization of the cost function, we use a Trust-Region method~\cite{conn2000TrustRegion}.
\begin{table*}[!tb]
    \centering    
    \begin{tabular}{c|c|c|c|c|c|c|c|c|c|c}
    	& \multicolumn{5}{c|}{$N=2$}& \multicolumn{5}{c}{$N=3$}\\
    	\hline
    	$|$State$\rangle$ & $|\phi_0\rangle$ & $|\psi_{P}^{(8)}\rangle$ & $|\psi_S^{(8)}\rangle$ & $|\psi_O^{(8)}\rangle$ & $|\phi_0(\Delta\mu_8^\text{o})\rangle$ & $|\phi_0\rangle$ & $|\psi_{P}^{(8)}\rangle$ & $|\psi_S^{(8)}\rangle$ & $|\psi_O^{(8)}\rangle$ & $|\phi_0(\Delta\mu_8^\text{o})\rangle$ \\
    	\hline
    	\hline
    	$\xi_C$ & 0.29515 & 0.20705 & 0.20905 & 0.20805 & 0.19249 & 0.18699 & 0.16724 & 0.16725 &0.16724 & 0.16421 \\
    	$\xi_N$ & & 0.21539 & 0.21539 & 0.21538 & 0.19272 & & 0.16729 & 0.16730 & 0.16740 & 0.16424 \\
    	$\xi_N$ [Eq.~\eqref{eq:xiN-theory}] &  & \multicolumn{3}{c|}{0.21715} & 0.19124 & & \multicolumn{3}{c|}{0.16690} & 0.16389\\
    	$R$ &2.702929  
        & 0.140104
        & 0.140105 
        & 0.139488 
        & 0.109123
        & 2.551906
        & 0.071830 
        & 0.071830 
        & 0.071823
        & 0.068051
        \\
    	$\mathcal{I}\left[|\psi_{P}^{(8)}\rangle\right]$ & & 0 &$6.8\times10^{-4}$  & $1.7\times10^{-4}$ & $4.9\times10^{-3}$ & & 0 & $4.4\times10^{-6}$ &$1.1\times10^{-6}$ & $1.0\times10^{-4}$\\
    	$\mathcal{I}\left[|\psi_S^{(8)}\rangle\right]$ & & $6.8\times10^{-4}$ & 0 & $1.7\times10^{-4}$ & $3.1\times10^{-3}$ & & $4.4\times10^{-6}$ & 0 &$1.1\times10^{-6}$ & $7.4\times10^{-5}$\\
    	$\mathcal{I}\left[|\psi_O^{(8)}\rangle\right]$ & & $1.7\times10^{-4}$ & $1.7\times10^{-4}$ & 0 & $3.8\times10^{-3}$ & & $1.1\times10^{-6}$ & $1.1\times10^{-6}$ & 0 & $8.6\times10^{-5}$\\
    \end{tabular}
    
    \caption{The correlation length $\xi_C$ (fit to numerics), density decay length $\xi_N$ [fit and theoretical approximation in \cref{eq:xiN-theory}], width $R$ [Eq.~\eqref{eq:R2}], and infidelities $\mathcal{I}\left[|\psi\rangle\right] = 1-|\langle \psi|\text{State}\rangle|^2$ for: ground state $|\phi_0\rangle$, projected boson stack $|\psi_{P}^{(8)}\rangle$, single-stack soliton $|\psi_{S}^{(8)}\rangle$, narrowest soliton $|\psi_{O}^{(8)}\rangle$, and ground state $|\phi_0(\Delta\mu_8^\text{o})\rangle$ with optimal pinning $\Delta\mu_8^\text{o}\simeq-3.66J$ ($-1.13J$) for $N=2$ ($N=3$) [$|\phi_0(\Delta\mu_8^\text{o})\rangle$ has the lowest infidelity with $|\psi_P^{(8)}\rangle$; the theoretical $\xi_N$ with pinning is discussed in Appendix~\ref{app:variational}]. See text for details about the evaluation of these quantities. Other parameters for the BH model are $U=-10J$ and $M=15$.
    }
    \label{tab:soliton-comparison}
\end{table*}

The states $|\psi_{P}^{(i)}\rangle$, $|\psi_S^{(i)}\rangle$ and $|\psi_O^{(i)}\rangle$ are generally different. In Tab.~\ref{tab:soliton-comparison} we compare the correlation length, decay length, width, and relative infidelities for these three quantum solitons (localized around the middle site $i_0=8$ in a $M=15$ sites chain) and the ground state of the BH model; for the latter, $\xi_N$ is not reported as the density does not decay exponentially [cf. Fig.~\ref{fig:BHm-eigenenergies-density}(d)]. The decay and correlation lengths are estimated by fitting the logarithm of the expectation values of $\hat{n}_i$ and $\hat{n}_i\hat{n}_j$, respectively (see Appendix~\ref{app:variational} for details). 

The density decay length $\xi_N$ is close to the correlation length $\xi_C$ for the three soliton states, all decaying faster (and hence being more localized) than what one would expect from the correlation length of the ground state $|\phi_0\rangle$. The density decay length for a spatially localized quantum soliton in the uniform system can be approximately computed for $|U|(N-1)\gg J\sqrt{N}$ through a variational procedure (see Appendix~\ref{app:variational}):
\begin{equation}
\label{eq:xiN-theory}
    \frac{1}{\xi_N}\simeq 2\ln\left[\frac{|U|(N-1)}{J}\right],
\end{equation}
which we use to plot the dashed lines in \cref{fig:BHm-eigenenergies-density}(d). We note that, according to Eq.~\eqref{eq:xiN-theory}, a localized soliton decays over a length smaller than the lattice spacing (unity in our notation) by a factor scaling with the inverse of the logarithm of $|U|(N-1)/J\gg 1$; hence, this state has no analog in the continuous limit.
For exponentially decaying density profiles, and assuming $\xi_N\ll M$, $R$ and $\xi_N$ are related by $R=1/\sqrt{2}\sinh (1/2\xi_N)$; in the regime of validity of Eq.~\eqref{eq:xiN-theory}, at leading order we recover the expression for $R$ derived in Ref.~\cite{Blain:2022qur} for a pinned soliton. The estimates of $\xi_N$ computed from the values of $R$ in Tab.~\ref{tab:soliton-comparison} according to this relation differ from the ones determined through fitting (c.f. Tab.~\ref{tab:soliton-comparison}) only by 0.1--0.2\%. 

Notably, the infidelities between the three different localized solitons are less than $10^{-3}$ for $N=2$ and of order $10^{-6}$ for $N=3$. Similarly, the length scales characterizing these states, $\xi_C$, $\xi_N$, and $R$, differ by a few parts per $10^{-3}$ for $N=2$ and per $10^{-5}$ for $N=3$. Thus, we can  conclude that the three protocols are all effective in defining a spatially localized packet made of soliton band states. For concreteness, in the rest of this work we will consider the projected boson stack $|\psi_{P}^{(i)}\rangle$. 

\section{Preparing quantum solitons in transmon arrays}
\label{sec:TransmonSoliton}

In the previous section, we identified the spatially localized quantum solitons in the BH model. In this section, we suggest a chain of capacitively coupled transmon qubits as a platform in which solitonic states can be implemented and manipulated. 

The Hamiltonian of $M$ capacitively coupled transmons
\begin{equation}\label{eq:arrayH}
    \mathcal{\hat H} = \sum_{i=1}^M
    4 E_C \left(\frac{\hat Q_i}{2e}\right)^2 - E_{J i} \cos \hat{\varphi}_i
    + \frac{T}{4e^2}\sum_{i=1}^{M-1}\hat Q_i\hat Q_{i+1}\,,
\end{equation}
where $\hat{\varphi}_i$ and $\hat{Q}_i$ are the phase and excess charge operators of transmon $i$ (canonically conjugated $[\hat{\varphi}_j,\hat{Q}_k]=2ei\delta_{jk}$ with $e$ the electron's charge), is approximately described by the BH Hamiltonian of Eq.~\eqref{eq:BHm} in which the bosonic degrees of freedom emerge as localized plasmonic excitations (see, e.g., Refs.~\cite{hacohen2015cooling,orell_probing_2019,mansikkamaki_beyond_2022,berke_transmon_2022}); in the transmon regime~\cite{koch_charge_2007}, the Josephson energy $E_{Ji}$ of each transmon is large compared to its charging energy $E_C$ (assumed, for concreteness, uniform). More precisely, at the first sub-leading order in $E_{Ji}/E_C\gg 1$, the transition energy between the ground and the first excited state in each transmon corresponds to the local chemical potential, $\mu_i = \omega_{01i}\simeq(\sqrt{8E_{Ji}E_C}-E_C)/\hbar$, and the charging energy gives the attractive interaction, $U \simeq -E_C/\hbar$. Finally, the hopping coefficients between neighboring sites are proportional to the capacitively coupling energy $T$, $J_{i,i+1} \simeq T\sqrt[4]{E_{Ji}E_{Ji+1}}/4\hbar\sqrt{2 E_C}$. The Josephson energies can be tuned on chip by replacing each Josephson junction by a SQUID and applying magnetic flux to each SQUID through dedicated flux lines~\cite{krantz_quantum_2019};
consequently the transmon frequencies (and so the chemical potentials) can be dynamically controlled~\footnote{This modulation affects, although typically in a reduced way, also the hopping coefficients.}. For typical transmons, with $E_J/E_C\simeq50$-$100$ and $E_C/h\simeq 100$-$300\,$MHz~\cite{koch_charge_2007,krantz_quantum_2019,Blais:2020wjs}, the chemical potential modulation $\Delta\mu$ can be as large as a few GHz, and so of the order of tens of $|U|$ in the Bose-Hubbard description, the latter quantity being also large compared to the hopping coefficient with typical values $J\simeq1$-$40\,$MHz~\cite{houck2012chip,berke_transmon_2022}. Note that the hopping coefficient can also be engineered to be sizably larger, see for instance  Ref.~\cite{riccardi2026experimentalobservationdynamicalblockade}, where $J/U\sim 0.6$-$0.8$ was reported in a device designed to maximize the ZZ coupling between qubits; in this case $|U|$ is comparable to $U_C$ for $N=3$. 

Below, we investigate a protocol to prepare a quantum soliton with high fidelity. Unless explicitly mentioned, in the numerics of this section we always consider a chain with $M=11$ sites. This array size satisfies two conditions: firstly, $M\gg \xi_N$ ensures that finite-size effects can be ignored for solitons localized near the array center;  secondly, $M \lesssim T_1\tilde{J}$, with the typical relaxation time $T_1 \gtrsim 30\,\mu$s for the first few excitation levels~\cite{Peterer2015Coherence,wang2025observingtwoparticlecorrelationdynamics}, makes it possible to study unitary evolution up to times for which at any site in the array a non-negligible probability to find an excitation has been reached while discarding the effect of relaxation. In the following, we first identify an optimal pinning strength value, and the drawbacks of an adiabatic soliton preparation protocol based on it, see Sec.~\ref{sec:optimal}. Then, in Sec.~\ref{sec:preparation}, we propose a non-adiabatic protocol to achieve high fidelity with a projected boson stack on a timescale substantially shorter than that governing the evolution of the typical multi-boson states.

\begin{figure}
    \centering
    \includegraphics[width=0.8\linewidth]{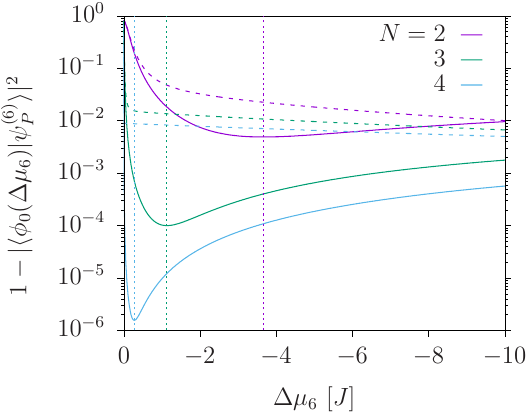}
    \caption{The infidelity of a projected boson stack (solid lines) and a boson stack (dashed lines) with the ground state of the BH Hamiltonian \cref{eq:BHm} as a function of the pinning $\Delta\mu_6$ in the central site for $M=11$ and $U=-10J$. The dotted vertical lines identify the pinning at which minimum infidelity is found for the projected stack for each $N$. 
    }
    \label{fig:fidelity-GS-stack}
\end{figure}
\subsection{Optimal pinning and drawbacks of adiabatic state preparation of a localized soliton}
\label{sec:optimal}
The wide tunability of the chemical potentials enables preparing localized states. As we noted in Sec.~\ref{sec:spatiallyLocalized}, while 
localization in a site $i_0$ can be achieved by pinning, the resulting state has finite occupation of the eigenstates of the uniform BH model outside the soliton band. Here we show that the pinning strength can be properly tuned to minimize the infidelity between the ground state of the Hamiltonian with pinning and the projected boson stack identified in Sec.~\ref{sec:spatiallyLocalized}.

\Cref{fig:fidelity-GS-stack} shows the infidelity of a projected boson stack with the ground state of the BH Hamiltonian with pinning $\Delta\mu_6=\mu_6-\mu_0$ in the central site for $M=11$ and $N=2,\,3,\,4$, where $\mu_0$ is the chemical potential of all the other sites. In all cases, the infidelity is a non-monotonic function of the pinning strength, and is minimized at an optimal value (see vertical dotted lines in Fig.~\ref{fig:fidelity-GS-stack}) that depends on the number of bosons. With increasing $N$, the infidelity decreases at a given pinning and the absolute value of the optimal pinning decreases. This non-monotonic behavior can be intuitively understood as follows. For very small pinning, $N \Delta \mu  \lesssim \tilde{J}$, the ground state is generally delocalized [see Fig.~\ref{fig:BHm-eigenenergies-density}(d) for the density distribution at zero pinning]. As the pinning strength increases, the ground state becomes progressively more localized and closer to a boson stack in the limit of infinitely large pinning -- see the dashed curves in \cref{fig:fidelity-GS-stack} showing the infidelity with a boson stack. Hence, there is an optimal value of the pinning for which the density localization is of the order of the one dictated  by the interaction only and, as for the projected boson stack, can be  characterized by the localization length $\xi_N$. At such optimal pinning, the ground state is in fact more localized than the projected boson stack due to the additional localization in the presence of a finite pinning, see \cref{app:variational}. Moreover, its infidelity with the localized states discussed in Sec.~\ref{sec:spatiallyLocalized} is much larger than those among the states themselves (cf. Tab.~\ref{tab:soliton-comparison}). While this result indicates that the pinned soliton is a distinct state, the magnitude of the infidelity is still small, so that it is still meaningful to prepare the optimally pinned soliton as a good proxy for the localized one. 
\begin{figure}
    \centering
    \includegraphics[width=\linewidth]{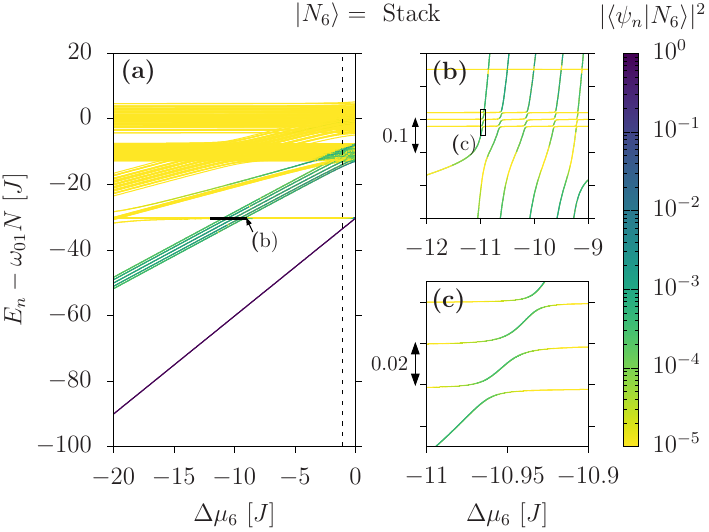}
    \caption{Energy spectrum of the BH Hamiltonian as a function of central pinning strength $\Delta\mu_6$ for $N=3$ bosons, $M=11$ sites, and $U=-10J$. The line color shows the overlap of each eigenstate with an $N$-particle boson stack on the central site. The dashed vertical line identifies the value of $\Delta\mu_6$ for optimal ground state fidelity with a projected boson stack. (b)-(c) Enlarged plots of the spectrum for energies $E_n-\omega_{01}N\sim -NU$ and pinning $\Delta\mu_6\sim U$, for which the states $\ket{(N-1)_6,1_{i\neq6}}$ are quasi-degenerate with the boson stacks $\ket{N_{i\neq6}}$ (see text).
    }
    \label{fig:spectrum-fidelity-stack}
\end{figure}

In contrast to cold-atom platforms (in which the number of bosons can be
large), in superconducting circuits, due to unavoidable relaxation processes, achieving the ground state of the pinned Hamiltonian with a fixed number of bosons is challenging. Stabilizing a state with a given number of bosons generally requires a specifically engineered driven-dissipative setup~\cite{Marcos2012PhotonCondensation,hacohen2015cooling,Ma2019DissipativeMottPhotons,schamrib2025refrigeration1dgasmicrowave}. Here we propose a protocol that circumvents this need.

For transmon arrays, Fock states of the BH model, such as boson stacks, can be prepared by applying suitable microwave pulses to the individual transmons. 
Since the boson stack corresponds asymptotically 
to the ground state of the Hamiltonian with an infinitely large pinning, we can, in principle, attempt to prepare the state through an adiabatic protocol in which we first initialize a boson stack in a strongly pinned site, and then ramp up the chemical potential adiabatically to the optimal value. 
\Cref{fig:spectrum-fidelity-stack} shows the spectrum of the BH model as a function of $\Delta\mu_6$ for $N=3$; the line color shows the square modulus of the coefficients of the spectral decomposition for a boson stack prepared in the central site. As the pinning $\Delta\mu_6$ is increased (in absolute value) from zero, the ground-state energy rapidly detaches from the soliton band by an energy $\sim N \Delta\mu_6$ and becomes the state with largest overlap with the boson stack in the pinning site. 
The other states in the soliton band are insensitive to the pinning strength. Of the states outside the soliton band, those with large occupation at the pinning site are also affected by the pinning; in particular, the energy of those states with $N-1$ bosons in the pinning site and one boson elsewhere, $\ket{(N-1)_6,1_{i\neq6}}$, decreases as $(N-1)\Delta\mu_6$ and crosses the soliton band at $\Delta\mu_6 \simeq U$. Note that 
a localized soliton consists mostly of a stack, $\ket{N_{i_0}}$, and states of the form $\ket{(N-1)_{i_0},1_{i_0\pm 1}}$ - see \cref{fig:spectrum-fidelity-stack}. Hence, for an adiabatic protocol to work, we need the energy of the states $\ket{(N-1)_{i_0},1_{i_0\pm 1}}$ 
to be below the soliton band, $|\Delta\mu_6| > U$. To reach the optimal pinning value [see dashed vertical line in Fig.\ref{fig:spectrum-fidelity-stack}(a)] starting from a large pinning (in absolute value)  $|\Delta\mu|> U$, it is therefore necessary to cross a region where the level spacings are much smaller than the coupling strength $J$, since they are smaller than the level spacings in the soliton band $\sim \pi\tilde{J}/(M+3-2\lceil N/2\rceil)$ [see Figs.~\ref{fig:spectrum-fidelity-stack}(b) and (c)];  the rate of change in
the chemical potential while passing this region would need to be slowed proportionally to the inverse of the level splittings for the process to be adiabatic (in other words, the duration of the adiabatic evolution must be much longer than $\sim M/\tilde{J}$, but the latter quantity is of order $T_1$, as discussed at the beginning of this section). For this reason, this protocol is not appealing on the practical side, given also the reduced lifetime of boson stacks compared to the single-boson states~\cite{Peterer2015Coherence,WangPhysRevApplied23}. We now introduce a different protocol to prepare a quantum soliton in a short time $\ll 1/J$.

\subsection{Non-adiabatic preparation of a localized soliton}\label{sec:preparation}
\begin{figure}
    \centering
    \includegraphics[width=\linewidth]{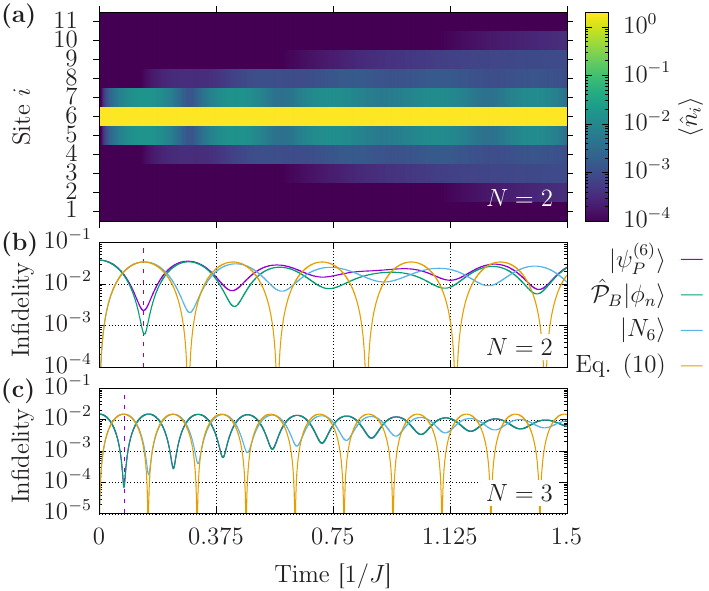}
    \caption{
    The constant pinning protocol for soliton preparation.
    \textbf{(a)} Time evolution starting from an $N=2$ boson stack with constant strong pinning for $\Delta\mu_6=-11.6J$. Panels \textbf{(b)} and \textbf{(c)} show the infidelity $\mathcal{I}\left[|\psi\rangle\right]$ of the evolved boson stack for $N=2$, $\Delta\mu_6=-11.6J$, and $N=3$, $\Delta\mu_6=-19.7J$, respectively, with the target quantum soliton in the uniform system (purple), the projector $\mathcal{\hat P}_B$ over the soliton band of the uniform system (green), and the initial state (blue). 
    In orange is the stack infidelity $1-|\langle N_6|\psi(t)\rangle|^2$ from \cref{eq:population-3state}.
    The vertical dashed purple lines show the time of the minima of target state infidelity.
    Data is for the BH model with $U=-10J$, $M=11$.}
    \label{fig:constant-dynamics}
\end{figure}

\begin{figure}
    \centering
    \includegraphics[width=\linewidth]{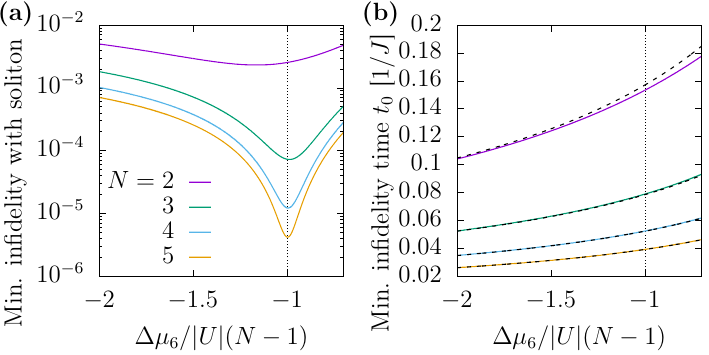}
    \caption{
    Finding the parameters for the constant pinning protocol.
    \textbf{(a)} Plot of the minimum infidelity with a quantum soliton in the uniform system for a boson stack evolved in the time range $0\ldots 1/J$ in a system with constant pinning $\Delta\mu_6$.
    \textbf{(b)} The time $t_0$ at which the minimum infidelity is taken. The dashed horizontal lines show $t_0^\text{opt}$ for each $N$.
    The dotted vertical line is at $\Delta\mu_6=U(N-1)$, intersections with data series giving the approximate value of $t_0^\text{opt}$. 
    Data is for the BH model with $U=-10J$, $M=11$.
    }
    \label{fig:constant-trajectory}
\end{figure}

Above, we commented on the main drawbacks of an adiabatic state preparation of a quantum soliton from a boson stack. Here, we investigate a protocol which can be implemented in a substantially shorter time, while still relying on the dynamic control of the chemical potential in the initialization site.
The boson stack $\ket{N_i}$ is directly coupled to the states $\ket{N-1_i \ 1_{i-1}}$ and $\ket{N-1_i \ 1_{i+1}}$ by the hopping term in the BH Hamiltonian, Eq.~\eqref{eq:BHm}. If we consider only time-scales up to $1/J$, we can approximate the unitary evolution of the boson stack restricting the analysis to the subspace generated by these three states. The evolution of a boson stack in this three-dimensional subspace can be computed explicitly by diagonalizing the restricted Hamiltonian and computing the Schr\"odinger evolution; the resulting probabilities read
\begin{align}\label{eq:population-3state}
|\braket{N_i|\psi(t)}|^2 &=1-\frac{2}{\lambda^2}\sin^2[\lambda t \sqrt{N}J ]\,,\\
\label{eq:population-next-state}
|\braket{N-1_i\ 1_{i\pm1}|\psi(t)}|^2 &=\frac{1}{\lambda^2}\sin^2[\lambda t \sqrt{N}J ]\,,
\end{align}
where 
\begin{equation}
\label{eq:lambda}
\lambda=\frac{1}{2}\sqrt{\frac{[(N-1)U+\Delta\mu_6]^2}{NJ^2}+8}\simeq-\frac{(N-1)U+\Delta\mu_6}{2\sqrt{N}J}\,,
\end{equation}
and the approximation in Eq.~\eqref{eq:lambda} holds for $-[(N-1)U+\Delta\mu_6]/{\sqrt{N}J}\gg 1$ -- see \cref{app:breathing} for details.

Equations~\eqref{eq:population-3state} and \eqref{eq:population-next-state} describe the population exchange between the boson stack and the states $\ket{N-1_i \ 1_{i\pm 1}}$. In this subspace, the time evolution is periodic with period 
\begin{equation}
T=\frac{\pi}{ J\lambda\sqrt{N}}\,.
\label{eq:t0optTheo}
\end{equation}
After a time $t_0=T/2$, a population fraction equal to $1/\lambda^2$ is transferred to each of the two states $\ket{N-1_i \ 1_{i+ 1}}$ and $\ket{N-1_i \ 1_{i- 1}}$. We can exploit this transfer to prepare a state with minimal infidelity with the projected stack. More precisely, we require this exchanged population to match the density of the sites adjacent to the initialization site in the projected stack. 

At the leading order in $|U|(N-1)/J\sqrt{N}\gg 1$ we find that the optimal value of the pinning is approximately (see \cref{app:breathing})
\begin{equation}
\label{eq:muOpt}
\mu^\text{opt} \simeq U(N-1)\, ,
\end{equation}
and the corresponding optimal evolution time to minimize the infidelity with the projected boson stack is therefore
\begin{equation}
t_0^\text{opt} \simeq \pi / 2 |U|(N-1)\, .
\end{equation}

\Cref{fig:constant-dynamics}(a) shows the evolution of the expectation value of the occupation numbers $\braket{\hat{n}_i(t)}$ of an $N=2$ boson stack in a system with $\Delta\mu_6=\mu^\text{opt}=-11.6J$. The occupation number of the pinned site decreases only slightly, while the adjacent sites are populated on the short timescale $t_0^\mathrm{opt}\ll 1/J$, in agreement with the analysis in the three-state subspace of \cref{app:breathing}; at longer times, the population `leaks' to sites farther from the pinning one. Figures~\ref{fig:constant-dynamics}(b) and (c) show the infidelity of the evolved state with: our target state, the projected boson stack (purple); its projection to the soliton band of the uniform system (green); and with the initial state, the boson stack (blue) for $N=2$ and $N=3$, respectively. The fidelity with the target soliton remains close to that with the soliton band projection, indicating that in the band the evolved state is largely the target one. The infidelity with the boson stack is captured by the evolution in the three-state subspace, \cref{eq:population-3state} (orange), for short time, capturing well both the period and maxima of the oscillations; for $N=3$, the agreement persists over longer times compared to $N=2$. Additional minima in the infidelity with the projected stack are evident around multiple (odd) integers of $t_0$, and they could also be exploited for state preparation: they are shallower, so their infidelity is higher compared to the first minimum, but less sensitive to timing errors. 

To benchmark our protocol, we compute the time evolution starting from the boson stack for $N=2,\ldots,5$ as a function of the pinning $\Delta\mu_6$, and identify the minimum infidelity over time with the corresponding projected boson stack, see \cref{fig:constant-trajectory}(a). Following our analysis in the reduced three-state subspace, in the plot we express the pinning strength in units of $|U|(N-1)$. Consistently with our analysis, the infidelity is non-monotonic with the pinning strength and is minimum for a value close to $\Delta\mu_6\approx U(N-1)$ (dashed vertical line), ranging from $99.8\%$ for $N=2$ to at least $99.999\%$ for $N=5$. Numerically, we find that the optimal value slightly deviates from the analytical prediction for small particle numbers; for instance, for $N=2$, we numerically find $\Delta\mu_6=-11.6J$, while Eq.~\eqref{eq:muOpt} returns $-10J$.

\Cref{fig:constant-trajectory}(b) shows the time $t_0$ corresponding to the minimum infidelity displayed in \cref{fig:constant-trajectory}(a). The solid lines are obtained numerically by computing the unitary evolution of the boson stack, while the dashed lines are given by Eq.~\eqref{eq:t0optTheo}. The time $t_0$ decreases with increasing $N$ and $|\Delta\mu_6|$, in agreement with our theoretical approximation. We note that, despite the clear minima around $\Delta\mu_6\approx U(N-1)$ in \cref{fig:constant-trajectory}(a), the time for minimum infidelity decreases monotonically for stronger pinning. This is because a stronger pinning both decreases the period of the oscillation of the stack state's occupation and decreases the decay length of the evolved state at the half-period of said oscillation -- cf. \cref{eq:population-3state,eq:population-next-state}, and \cref{app:breathing}. 

We close this section remarking that the projected boson stack in the uniform system can be obtained with the fidelity discussed above by quenching the chemical potential $\mu^\text{opt}\to 0$ in the initialization site at time $t_0$ after the unitary evolution. In Appendix~\ref{sec:preparation-dynamics} we discuss how the protocol can be adjusted, with no appreciable decrease of fidelity, when considering a more realistic finite-time quench.

\section{Quantum walk of a spatially localized soliton }\label{sec:single-particle}

A defining feature of classical solitons is that they retain their shape during propagation~\cite{Russell1844Report}. This feature does not necessarily hold for quantum solitons due to the wave-like properties of the quantum evolution. Since a spatially localized soliton is not an eigenstate of the BH model, it evolves unitarily and spatial localization is generally lost in time. In this section, we investigate the unitary evolution of localized solitons along with that of a single particle, projected boson stacks, and the state prepared with our protocol of Appendix~\ref{sec:preparation-dynamics}.

To highlight the role of interaction, we first review the evolution of a single boson placed initially on site $i$. The occupation number of site $j$ becomes finite on a timescale proportional to $J$, due to hopping, and inversely proportional to the distance to the initial site $|i-j|$. The unitary time evolution of the initial state manifests itself in a specific interference pattern known as quantum walk.
As already discussed in Sec.~\ref{sec:quantumSolBH}, boson stacks behave as composite particles with effective hopping amplitude $\tilde J$ for strong attractive interaction $-U\gg J$~\cite{mansikkamaki_beyond_2022}. Thus, a boson stack in site $i$ approximately performs a quantum walk around the site with an ``expansion velocity'' of order $\sqrt{2}\tilde{J}$, and so does the ground state of the BH model with a pinning potential $\mu_i<0$ after removing the pinning~\cite{Blain:2022qur}. 
In the latter work,
we argued that, compared to a boson stack, the ground state with pinning
is substantially less affected by residual single-particle components, i.e., states outside the soliton band, propagating at speed $J$. Here, we make a more quantitative investigation of this point, comparing boson stacks and projected boson stacks (that is, localized quantum solitons), the latter being a superposition of states in the band $B$ only. 

\begin{figure*}
    \centering
    \includegraphics[width=\linewidth]{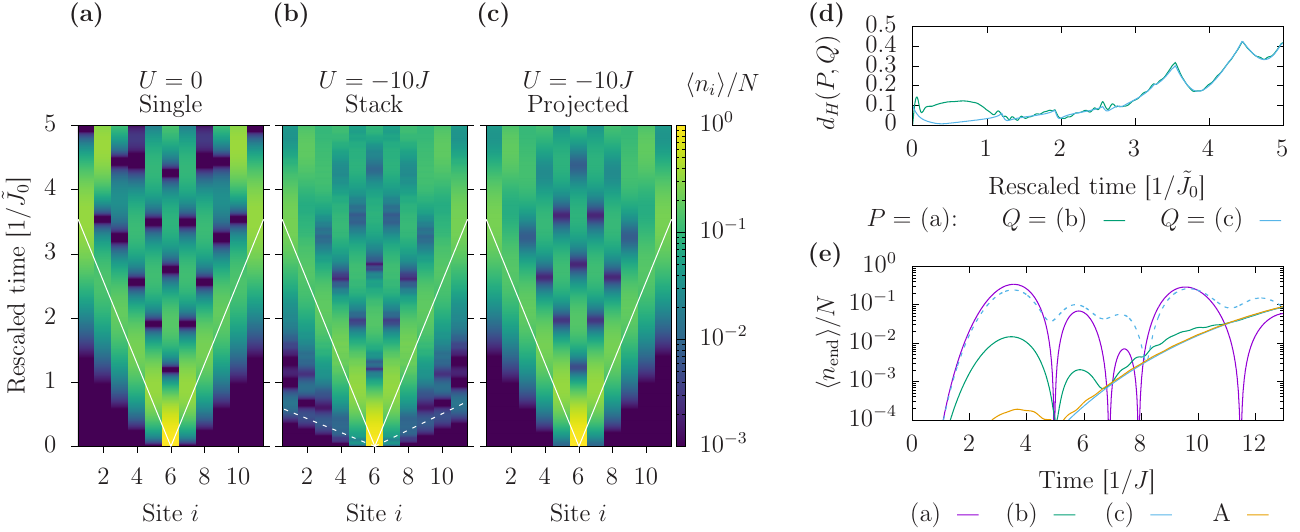}
    \caption{Per-site density evolution for (a) a single particle, (b) a boson stack and (c) a localized soliton (projected stack) created and evolved with $N=2$ bosons, $M=11$ sites, and $U=-10J$ for the interaction strength.  Time has been rescaled by $\tilde J_0 = J$ for $U=0$ and $\tilde J_0 = 0.198 J$ for $U=-10J$ -- see main text. The white lines indicate the edges of the quantum walk as predicted by the ``expansion velocity'' $\sqrt{2} \tilde{J}_0$ ($\sqrt{2} J$ for dashed). Panel (d) shows the Hellinger distance, \cref{eq:dH}, where $P$ and $Q$ are the density distributions for the three left panels as detailed in the legend. Panel (e) shows the density of the first 
    site with time for each of the panels. The data series marked ``A'' (orange curve) shows the evolution of the localized soliton prepared with our protocol -- see  main text. The dashed light blue curve shows the data for panel (c) with time rescaled by $\tilde J_0\approx \tilde J$.}
    \label{fig:single-particle-comparison}
\end{figure*}

\begin{figure*}
    \centering
    \includegraphics[width=\linewidth]{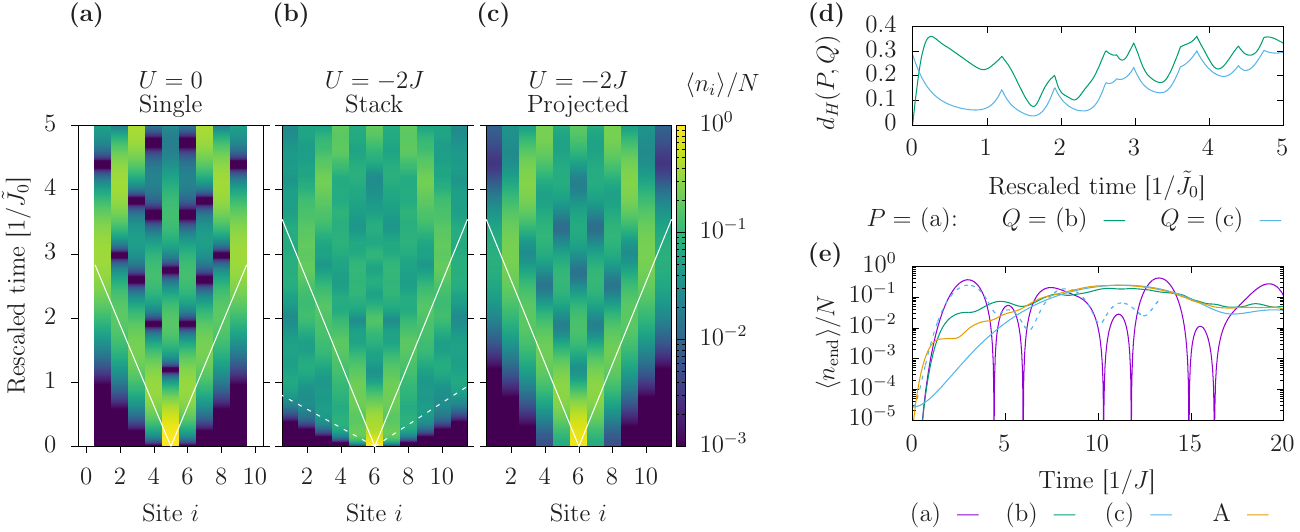}
    \caption{Same plots as in Fig.~\ref{fig:single-particle-comparison}, here with $N=3$ and $U=-2J$; for the finite-$U$ time rescaling we set $\tilde J_0 = 0.265 J$. In panel (a) we use $M-2$ sites; relatedly, in panel (e) $\hat{n}_\mathrm{end}$ denotes $\hat{n}_1$ for $U=0$ and $\hat{n}_2$ otherwise -- see main text.
    For curve A, the soliton is prepared as in~\cref{sec:preparation-dynamics} with parameters $t_0=0.3752/J$, $t_q=0.005/J$, and $\Delta\mu_6=-4.54J$.
    }
    \label{fig:single-particle-comparison-N3-lowU}
\end{figure*}

\Cref{fig:single-particle-comparison} shows the time evolution of the expectation value of the occupation numbers $\braket{\hat{n}_i}$ for a single boson [panel (a)], a boson stack [panel (b)], and a projected boson stack [panel (c)] initialized on site $i_0=6$ (with $N=2$ and $U=-10 J$). Due to the different expansion velocities, in the vertical axes we scale the time differently for panel (a), where $t$ is in units of $1/J$, compared to panels (b) and (c), where the time is in units of $1/\tilde{J}_0$ with $\tilde{J}_0 = 0.198 J$. The latter value is found by inspecting the time evolution the end-site occupation number for the single particle and the projected boson stack:  we set $\tilde{J}_0=J t_S/t_P$, where $t_S$ and $t_P$ are the shortest times at which $\braket{\hat{n}_1(t)}$ has a local maximum for a single particle and a projected stack, respectively. We find that $\tilde{J}_0$ differs from the analytical approximation $\tilde{J}$ [Eq.~\eqref{eq:Jtilde_def}] by $\sim1\%$  for the parameters as in \cref{fig:single-particle-comparison}.

The evolution of the three states is qualitatively similar, featuring an interference pattern which is typical of a quantum walk~\cite{karamlou_quantum_2022,wang2025observingtwoparticlecorrelationdynamics}. For the boson stack, \cref{fig:single-particle-comparison}(b), single particle features (at the subpercent level) propagating at speed $J$ are superimposed onto the collective motion. Such features are absent from the spatially localized soliton, \cref{fig:single-particle-comparison}(c).
To quantify the difference between the three state evolutions, 
we employ the Hellinger distance $d_\text{H}$, a metric used for comparing probability distributions which also finds wide applications for state tomography~\cite{PhysRevA.69.032106,science.1250147,PhysRevA.97.062342,Acharya_2019}. The Hellinger distance between two discrete probability distributions $P=(p_1, p_2,\ldots,p_M)$ and $Q=(q_1, q_2,\ldots,q_M)$ reads
\begin{equation}
    d_\text{H}(P,Q)=\frac{1}{\sqrt{2}}\sqrt{\sum_{i=1}^M\left(\sqrt{p_i} - \sqrt{q_i}\right)^2}\,.
    \label{eq:dH}
\end{equation}
In our analysis, the discrete components of each distribution are given by the normalized occupation numbers at time $t$, $\langle\hat{n}_i(t)\rangle/N$. To account for the different timescales, we compare the distributions at equal values of the scaled time $Jt$ for single particle and $\tilde{J}_0t$ for boson stack and quantum soliton. 
\Cref{fig:single-particle-comparison}(d) shows the Hellinger distance, \cref{eq:dH}, between the normalized occupation numbers 
for a single particle and the one for the boson stack (purple) and the soliton (green). Despite being larger at the initial time, the Hellinger distance between the soliton and the single-particle distributions becomes quickly substantially smaller than the one computed from the boson stack and remains smaller up to $t\simeq 1.2/\tilde{J}$. At larger times, the Hellinger distances are almost identical. Therefore, the short-time Hellinger distance could, in principle, be used to discriminate experimentally between a boson stack and a quantum soliton. 

As an alternative and experimentally less demanding approach to distinguish between boson stack and soliton we propose monitoring the occupation number at a single site sufficiently far from the initialization
site. This approach is based on the observation that the single-particle components of the boson stack propagate towards the end sites of the chain substantially faster than the soliton-band components. 
\Cref{fig:single-particle-comparison}(e) shows the expectation value at the edge site $\braket{\hat{n}_1}/N$ as a function of time (in units of $1/J$ for all the curves). For the  zero-interaction case [equivalent to a single particle, cf. Fig.~\ref{fig:single-particle-comparison} (a)], the boson arrives at the end-site after a time $
\sim(M-1)/2\sqrt{2}J$ and the normalized occupation exceeds values of order $1/M$.
For the boson stack, the time evolution is similar but the maximum value is over one order of magnitude lower. 
Finally, for the quantum soliton the bosons reach and populate the first site only after a substantially longer time.
The dashed curve gives $\braket{\hat{n}_1}/N$ for the soliton with time in units of $1/\tilde{J}_0$, showing that the end-site occupation closely follows that of the noninteracting case up to the time rescaling (we remind that the positions of the first maxima are aligned by our definition of $\tilde{J}_0$). The additional orange curve labeled ``A'' shows the evolution of $\braket{\hat{n}_1}/N$ for the state prepared following our protocol 
in~\cref{sec:preparation-dynamics}, for which the first maximum in the first-site density is two orders of magnitude less than for the boson stack. At longer times, the prepared state exhibits essentially the same evolution in first-site density as the quantum soliton in panel (c). 

As a second example of a quantum walk, we consider the case $N=3$ and $U=-2J$, the latter being an interaction strength slightly above (in absolute value) the critical value $U_C \simeq \sqrt{3}J$ at which the soliton band separates from the other states (cf. Sec.~\ref{sec:quantumSolBH}). Because of the weaker interaction, the soliton decay length is longer, $\xi_N \simeq 0.335$ (from fitting as in Sec.~\ref{sec:spatiallyLocalized}), compared to the $U=-10J$ case [cf. Table~\ref{tab:soliton-comparison}]. Correspondingly, the interference pattern for the soliton quantum walk  in Fig.~\ref{fig:single-particle-comparison-N3-lowU}(c) is less marked compared to that in Fig.~\ref{fig:single-particle-comparison}(c). However, the pattern for the corresponding stack, Fig.~~\ref{fig:single-particle-comparison-N3-lowU}(b), is even less clear. This qualitative assessment is quantitatively confirmed by the Hellinger distance plot in Fig.~\ref{fig:single-particle-comparison-N3-lowU}(d): at short (rescaled) times, both the soliton and stack distances are higher than in panel Fig.~\ref{fig:single-particle-comparison}(c); we note that due to the edge chemical potentials, the Hellinger distance here is calculated excluding the end sites (with distributions re-normalized accordingly) and taking as a reference the quantum walk at $U=0$ in an array of length $M-2$. Interestingly, at all times in the interval considered, except at very short time, the soliton Hellinger distance remains smaller than the stack distance.

Taking again into consideration the edge chemical potentials, in panel (e) we plot the normalized density in the first site for $U=0$ and in the second site for finite $U$. Again we find that it takes a longer time to achieve percent-level density for the soliton compared to the stack. This remains true also for the imperfect soliton prepared according to the protocol of~\cref{sec:preparation-dynamics}; we note that in this case the preparation time $\sim0.38/J$ is relatively long and the fidelity with the ``ideal'' localized soliton $\sim 93.6\%$ relatively low, due to the weakness of the interaction.

\subsection{Transmon eigenstate occupation}\label{sec:transmon-eigenstate-population}

\begin{figure}
    \centering
    \includegraphics[width=\linewidth]{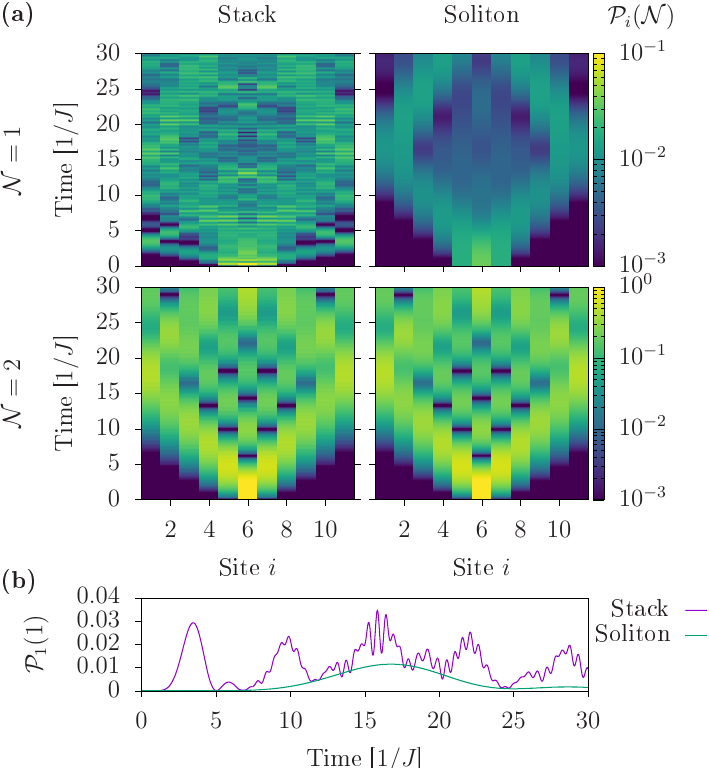}
    \caption{\textbf{(a)} Evolution of the probability $\mathcal{P}_i(\mathcal{N})$ of finding $\mathcal{N}=1$ (top row) and $\mathcal{N}=2$ (bottom row) bosons on site $i$ when starting with a boson stack (left) and a projected boson stack (soliton, right). 
    Panel \textbf{(b)} shows the Hellinger distance $d_H$ between the $\mathcal{N}=1$ and $\mathcal{N}=2$ probabilities for the stack and soliton.
    Panel \textbf{(c)} shows the evolution of the occupation of the first excited level of the end transmon for both the stack and soliton. 
    Data is for the BH model with $N=2$, $U=-10J$, and $M=11$.
    }
    \label{fig:projected-density}
\end{figure}

\begin{figure}
    \centering
    \includegraphics[width=\linewidth]{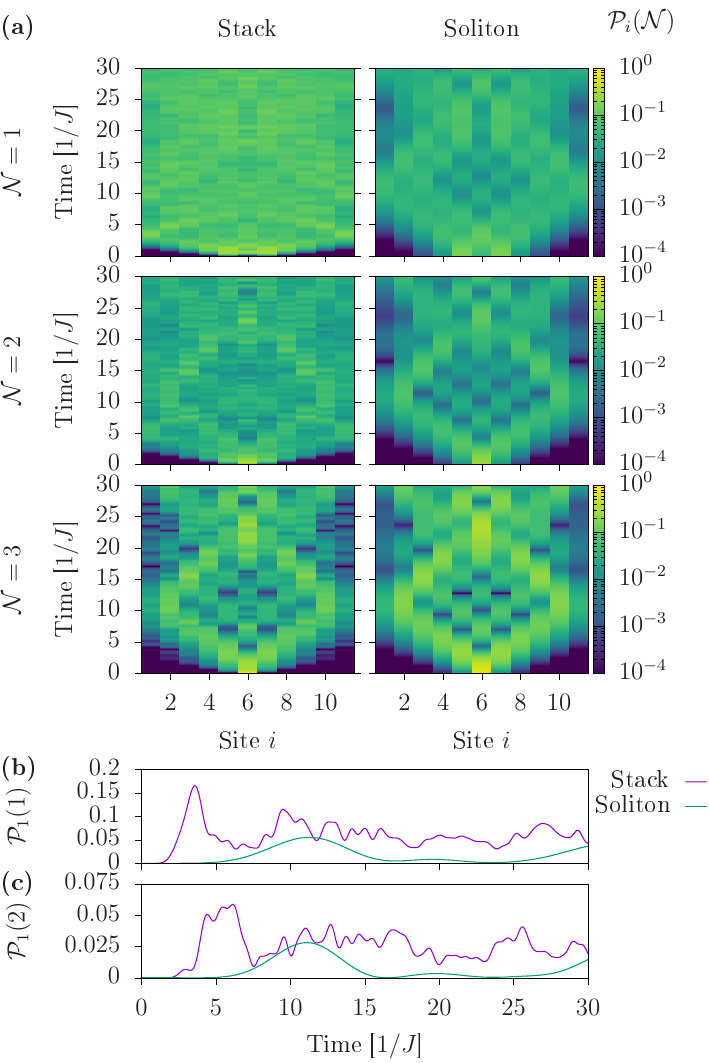}
    \caption{\textbf{(a)} Evolution of the probability of finding $\mathcal{N}=1$ (top row), $\mathcal{N}=2$ (middle row), and $\mathcal{N}=3$ (bottom row)  bosons on site $i$ when starting with a boson stack (left) and a soliton (projected boson stack, right). Panel \textbf{(b)} and \textbf{(c)} show the evolution of the occupation of the first and second excited level of the end site for both the stack and the soliton. 
    Data is for the BH model with $N=3$, $U=-2J$, and $M=11$.
    }
    \label{fig:projected-density-N3-lowU}
\end{figure}

Above, we discussed two possible observables to discriminate the boson stack from the quantum soliton based on the expectation values of the occupation numbers. More information can be obtained by extending the analysis to all the possible outcomes for the occupation numbers. In fact, in a single-shot readout of the superconducting qubits, the first few levels of each transmon, $\ket{0}$, $\ket{1}$, $\ket{2}$, \ldots, can be discriminated~\cite{BianchettiPRL105,chen2023transmon,WangPhysRevApplied23}. This enables studying the time evolution of the individual probabilities $\mathcal{P}_i(\mathcal{N})$ of finding $\mathcal{N}$-bosons on site $i$; these probabilities satisfy the normalization conditions
$\sum_{\mathcal{N}=0}^N \mathcal{P}_i(\mathcal{N})=1$ and $\sum_{i=1}^M\sum_{\mathcal{N}=1}^N \mathcal{N}\mathcal{P}_i(\mathcal{N})=N$, the latter expressing the conservation of the total boson number.

\Cref{fig:projected-density}(a) shows the time evolution of $\mathcal{P}_i(\mathcal{N})$ for $\mathcal{N}=1,2$ and $N=2$ total bosons, comparing an initially prepared boson stack (left panels) with a projected stack (right panels). The probabilities for the second excited state (bottom panels) evolve similarly for the two initial states and closely resemble the evolution of the occupation numbers displayed in Fig.~\ref{fig:single-particle-comparison}(b) and (c), respectively. Notably, more striking differences emerge when considering the evolutions of the single boson probabilities $\mathcal{P}_i(1)$.
In particular, the edges of the interference pattern for the soliton resemble those for $\mathcal{P}_i(2)$ as well as those for the noninteracting case seen in \cref{fig:single-particle-comparison}(a), showing that the two bosons comprising the soliton behave as a composite particle. 
In contrast, $\mathcal{P}_i(1)$ 
for the stack exhibits additional fast-moving components arising from eigenstates outside the soliton band. 

In analogy to \cref{fig:single-particle-comparison}(e), in \cref{fig:projected-density}(c) we show the time evolution of the probability of having a single boson in the end-site transmon. While for the soliton a single peak emerges at $t\simeq (M-1)/2\sqrt{2}\tilde{J}_0 \simeq 17/J$, indicating that all the bosons are moving together, the stack displays a first peak around the time $\sim (M-1)/2\sqrt{2}J$ characteristic of single-particle hopping. We conclude that monitoring the probability of finding the end-site transmon in the first excited state is a suitable route for discriminating a spatially localized soliton from a boson stack. 

The considerations above hold also for our second example ($N=3$, $U=-2J$). In this case the interference patterns are clearly more prominent for the soliton compared to the stack, in particular for the higher levels, as $P_i(3)$ and $P_i(2)$ more strongly resemble each other -- see Fig.~\ref{fig:projected-density-N3-lowU}(a). The ``composite particle'' nature of the soliton is also manifested when comparing the first-site probabilities $P_1(1)$ and $P_1(2)$ in Figs.~\ref{fig:projected-density-N3-lowU}(b) and (c), which evolve in a very similar way for the soliton. In contrast, for the stack $P_1(1)$ surges earlier than $P_1(2)$, and both earlier than the corresponding soliton probabilities, similarly to Fig.~\ref{fig:projected-density}(b).

\section{Conclusions}\label{sec:conclusions}

In this work, we studied the evolution of spatially localized quantum solitons in the Bose-Hubbard model.
While applicable to general arrays simulating the Bose-Hubbard Hamiltonian, we tailor part of our analysis -- the soliton preparation -- specifically to transmon platforms. 

In an array of size $M$, the $M$ lowest-energy eigenstates of the BH Hamiltionian are superpositions of spatially localized quantum solitons~\cite{naldesi_rise_2019}. In Sec.~\ref{sec:spatiallyLocalized} we numerically construct spatially localized solitons as superpositions of these solitonic eigenstates in three different ways, obtaining consistent physical features -- see Table~\ref{tab:soliton-comparison}. In Sec.~\ref{sec:TransmonSoliton} we explore protocols to prepare spatially localized quantum solitons in arrays of transmons. For typical experimental parameters, we show that high-fidelity preparation is possible, even when taking into account the finite rate at which the qubit frequency can be tuned by flux. Alternative preparation schemes are possible in other platforms, for instance in cold-atom settings where reaching ground states with optimal pinning strength [cf. Fig.~\ref{fig:fidelity-GS-stack}] can be experimentally feasible.

In \cref{sec:single-particle}, we quantitatively compare the quantum walk of a quantum soliton to that of a single particle. We show that the quantum soliton evolution demonstrates composite-particle behavior, both in the average occupation number [Figs.~\ref{fig:single-particle-comparison} and \ref{fig:single-particle-comparison-N3-lowU}] as well as in the probability to find a given number of bosons in each site of the array [Figs.~\ref{fig:projected-density} and \ref{fig:projected-density-N3-lowU}]. We also propose monitoring sites near the array edge as a way to discriminate between solitons and other states, in particular boson stacks.

While our work addresses fundamental properties of small (low number of bosons) localized quantum solitons, we point out that quantum sensing applications of such states could be envisioned for enhanced measurements of rotation and magnetic fields~\cite{atom_roadmap}.

\acknowledgments

BB would like to thank Gabrielle Roberts for discussions about their adiabatic melting technique for single excitations.

\appendix

\section{Fitting of, and analytical approximation for the soliton decay length}
\label{app:variational}

In Sec.~\ref{sec:spatiallyLocalized} we discuss three different procedures -- with overall equivalent results -- to identify spatially localized solitons. In this appendix, we provide details about the fitting procedure for determining $\xi_N$ and $\xi_C$ from the numerics, and the derivation of the analytical approximation for the decay length $\xi_N$, Eq.~\eqref{eq:xiN-theory}. 

As mentioned in the main text, we estimate decay and correlation lengths by fitting the logarithm of the expectation values of $\hat{n}_i$ and $\hat{n}_i\hat{n}_j$; more precisely, we perform a linear fit $y_\beta=\alpha_\beta-|i-i_0|/\xi_\beta$ ($\beta=\{N,C\}$) where $\alpha_\beta$ is approximately the expectation value at site $i_0$ and $\xi_\beta$ the decay (N) or correlation (C) length. In fitting, we exclude the central site $i_0=8$, due to the deviations from the exponential behavior in proximity of the localization site, and the edge sites $i=1$ and $i=M$, where the effective chemical potential $\tilde{\mu}$ is maximum (we note that excluding an additional pair of sites from the fit affects the values reported in Tab.~\ref{tab:soliton-comparison} for $\xi_C$ and $\xi_N$, up to the last two digits for $N=2$ and up to the last digit for $N=3$).

To obtain an analytical approximation of $\xi_N$, we focus on the strong interaction limit (in the sense specified below),
we consider the following ansatz for a soliton localized at site $i$,
\begin{equation}
\label{eq:ansatz}
\ket{\psi_\beta^{(i)}}=\sqrt{1-2\beta^2}\ket{N_i}+\beta(|N-1_{i}\,1_{i-1}\rangle+|N-1_{i}\,1_{i+1}\rangle)\,,
\end{equation}
where in addition to the boson stack at site $i$ we only include states which are coupled to it by a single boson hopping [the expression $\ket{N_{i},N_{j}} = \left(\hat{b}^\dagger_i\right)^{N_i}\left(\hat{b}^\dagger_j\right)^{N_j} \ket{0}/\sqrt{N_i!N_j!}$ denotes a state with $N_i$ (and $N_j$) bosons at site $i$ (and $j$) and no bosons in the remaining sites]. Moreover, we assume that the state should be symmetric around the site $i$, and so it can be expressed in terms of a single parameter ($\beta$). This parameter can be found by energy minimization, namely, we consider the expectation value of the Bose Hubbard Hamiltonian for the state in Eq.~\eqref{eq:ansatz}
\begin{align}
f^{(i)}(\beta)&=\braket{\psi_\beta^{(i)}|\frac{\hat H_{BH}}{\hbar}|\psi_\beta^{(i)}}\nonumber\\
&=(1-2\beta^2)\frac{U}{2}N(N-1)+2\beta^2\frac{U}{2}(N-1)(N-2)\nonumber\\
&+4\beta J\sqrt{N(1-2\beta^2)}
\label{eq:fiofBeta}
\end{align}
Requiring the condition for a stationary point, i.e., $df/d\beta=0$, and linearizing the resulting equation for $\beta\ll 1$, we find
\begin{equation}\label{eq:beta}
\beta=\frac{J\sqrt{N}}{U(N-1)}\,,
\end{equation}
a result which is in agreement with Ref.~\cite{BuonsantePRA72} (see Appendix B in this reference); the smallness of $\beta$ defines the strong interaction regime for the purposes of this appendix.
We can estimate the decay length by computing the ratio between the number occupation in the site $i$ and the one adjacent to it:
\begin{equation}
e^{-1/\xi_N}=\frac{\braket{\psi_\beta|\hat{n}_{i-1}|\psi_\beta}}{\braket{\psi_\beta|\hat{n}_{i}|\psi_\beta}}\simeq\frac{\beta^2}{N}
\label{eq:conditionxi}
\end{equation}
where the approximation in this equation holds at the leading order in $\beta\ll1$. Equation Eq.~\eqref{eq:conditionxi}
immediately returns
\begin{equation}
\frac{1}{\xi_N}=\ln\left[\frac{U^2(N-1)^2}{J^2}\right]\,,
\end{equation}
as reported in Sec.~\ref{sec:spatiallyLocalized}.
The ansatz in Eq.~\eqref{eq:ansatz} cannot be used at the edge sites $i=1,M$. For a soliton spatially localized at the (left) edge, the ansatz is modified as follows:
\begin{equation}
\ket{\psi_\beta^{(1)}}=\sqrt{1-\beta^2}\ket{N_1}+\beta|N-1_{1}\,1_{2}\rangle,
\end{equation}
where $\beta$ can be computed minimizing the energy function $f^{(1)}(\beta)$. After some algebra, we find that $f^{(1)}(\beta)$ can be obtained replacing $2\beta^2\to \beta^2$ and $4\beta\to 2\beta$ in the second line and the third line of Eq.~\eqref{eq:fiofBeta}. After energy minimization, one obtains again Eq.~\eqref{eq:beta}.
Inserting Eq.~\eqref{eq:beta} in $f^{(i)}(\beta)$ and $f^{(1)}(\beta)$, we can estimate the energy difference between a soliton localized at an edge site compared to an internal site, 
\begin{equation}
    f^{(1)}(\beta)-f^{(i)}(\beta)=\frac{J^2 N}{|U|(N-1)}\,.
\end{equation}
Notably, this result corresponds to the effective chemical potential difference between the edge and its adjacent site,
as reported in Ref.~\cite{mansikkamaki_beyond_2022} and quoted in Sec.~\ref{sec:quantumSolBH}.

The above approach can be modified to account for pinning of strength $\Delta\mu$ at site $i$; in practice, the pinning can be incorporated by replacing $U(N-1)\to U(N-1)+\Delta\mu$ in the coefficient $\beta$ of Eq.~\eqref{eq:beta}, an approach which is valid for $|U(N-1)+\Delta\mu|\gg J \sqrt{N}$. Consequently, the correlation length can be expressed as
\begin{equation}
\label{eq:xiN-theory_pin}
    \frac{1}{\xi_N}\approx 2\ln\left[\frac{|\Delta\mu+U(N-1)|}{J}\right].
\end{equation}
This expression shows that in the non-interacting case $U=0$ a localized state can be obtained due to the pinning alone.

\section{Short-time evolution of a boson stack under pinning}\label{app:breathing}

In order to understand the evolution of the boson stack $\ket{N_i}$ on a short timescale $\lesssim 1/J$, we write the matrix representation of the Bose-Hubbard model in the subspace generated by $\ket{N_i}$ and the states which are connected to it by a single hopping process $\ket{N-1_{i}\ 1_{i\pm 1}}$, the same states used also to estimate the localized soliton decay length in Appendix~\ref{app:variational}.
Up to an irrelevant constant matrix, the truncated Hamiltonian in the ordered basis $\{\ket{N-1_{i}\ 1_{i- 1}},\ket{N_i},\ket{N-1_{i}\ 1_{i+ 1}}\}$  reads 
\begin{equation}
\frac{\hat{\mathcal{H}}_F}{\hbar} = J\sqrt{N}
  \begin{pmatrix}
    0 & 1 & 0\\
    1 & \epsilon & 1\\
    0 & 1 & 0
  \end{pmatrix},
  \label{mat:3x3}
\end{equation}
where $\epsilon=[U(N-1)+\mu_i]/J\sqrt{N}$ is the (scaled) energy detuning between the Fock states $\ket{N_i}$ and $\ket{N-1_{i}\ 1_{i\pm 1}}$ under a pinning $\mu_i$. The eigenstates and the eigenvalues of the matrix in Eq.~\eqref{mat:3x3} can be readily found, 
\begin{align}
E_n&=J\sqrt{N}\Big\{\frac{\e}{2}-\frac{\sqrt{\e^2+8}}{2},0,\frac{\e}{2}+\frac{\sqrt{\e^2+8}}{2}\Big\}\,,
\label{eq:energiesmat3x3}\\
\ket{\varphi_1}&=\frac{\{1,\frac{\e}{2}-\frac{\sqrt{\e^2+8}}{2},1\}}{\sqrt{2+\left(\frac{\e}{2}-\frac{\sqrt{\e^2+8}}{2}\right)^2}
}\,,
\label{eq:phi1mat3x3}\\
\ket{\varphi_2}&=\frac{1}{\sqrt{2}}\{1,0,-1\}
\, ,
\label{eq:phi2mat3x3}
\\
\ket{\varphi_3}&=\frac{\{1,\frac{\e}{2}+\frac{\sqrt{\e^2+8}}{2},1\}}{\sqrt{2+\left(\frac{\e}{2}+\frac{\sqrt{\e^2+8}}{2}\right)^2}\,
}\,.
\label{eq:phi3mat3x3}
\end{align}
The evolution of the boson stack $\ket{N_i}=\{0,1,0\}$ is then  computed
decomposing the unitary time evolution operator on the eigenstates basis, 
\begin{equation}
\ket{\psi}=\sum_{n=1}^3 e^{-iE_nt}\braket{\varphi_n|N_i}\ket{\varphi_n}.
\label{eq:statemat3x3}
\end{equation}
Using Eqs.~\eqref{eq:energiesmat3x3}-\eqref{eq:statemat3x3} and after some algebra, we obtain the probabilities of Eqs.~\eqref{eq:population-3state} and \eqref{eq:population-next-state}, where $\lambda=\sqrt{\e^2+8}/{2}$ [cf. Eq.~\eqref{eq:lambda}]. As discussed in the main text, for $t=\pi/2\lambda J\sqrt{N}$ a population which amounts to $1/\lambda^2$ is transferred from the boson stack to the states $\ket{N-1_{i}\, 1_{i\pm 1}}$. The minimum infidelity with the spatially localized soliton is found by equating this population to the corresponding population in the soliton state. Requiring that $1/\lambda^2=\beta^2$, and using the approximation in \cref{eq:beta}, we find 
\begin{align}
\mu_{\rm opt} &=-U(N-1) - 2 \sqrt{(N-1)^2U^2-2J^2N}\\
&\simeq U(N-1)\,.
\end{align} 
where the approximation in the second line is found keeping the leading order term for $(N-1)|U|/\sqrt{N}J\gg 1$.

The considerations above also guides us in the numerical study of the evolution of a boson stack under a time-dependent chemical potential $\mu_i(t)$ on a target site $i$, see Appendix~\ref{sec:preparation-dynamics}. To compute the time evolution, we use standard Trotterization~\cite{Avtandilyan2024Optimal}: the evolution is computed by discretizing the time interval in steps of duration $\delta t$, and the state at time $t+\delta t$ is computed from the state at time $t$ as
\begin{equation}
|\psi(t+\delta t)\rangle = \exp[-\imath \hat{\mathcal{H}}(t) \delta t/\hbar]|\psi(t)\rangle\, ,
\end{equation}
where $\hat{\mathcal{H}}(t)$ is the Hamiltonian at time $t$. To compute the exponential of the operator, we perform a spectral decomposition of
the instantaneous Hamiltonian via exact diagonalization. In the calculation shown in the main text, we used the time step $\delta t=10^{-4}/J$ which is short compared to $T$ of \eqref{eq:t0optTheo}. In fact, within numerical precision we observed no difference in the final state in the time explored even if increasing the time step to $\delta t=10^{-3}/J$. 

\section{
Quenching to the uniform system
}\label{sec:preparation-dynamics}

\begin{figure}
    \centering
    \includegraphics[width=\linewidth]{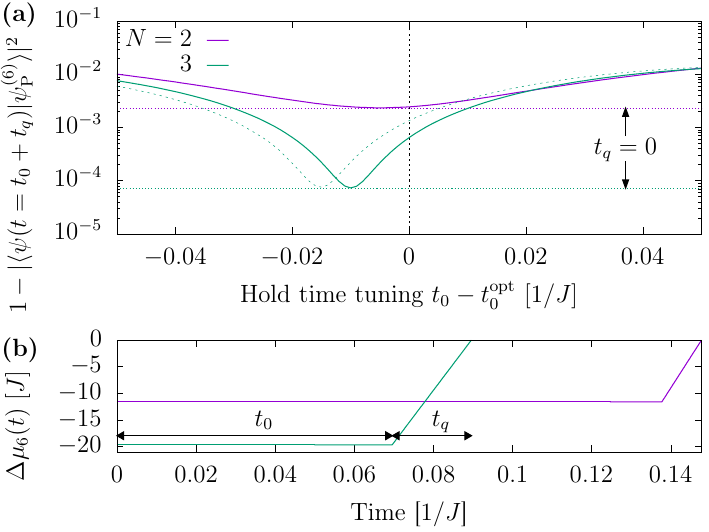}
    \caption{\textbf{(a)} Tuning of the hold time $t_0$ to minimize the infidelity with the localized quantum soliton in the uniform system $1-|\langle \psi(t=t_0+t_q)|\psi_\text{PF}\rangle|^2$. The horizontal axis shows the deviation $t_0-t_0^\text{opt}$ from the hold time for optimal infidelity without the quench. The dashed horizontal lines show the minimum soliton infidelity achieved with an instantaneous quench to the uniform system $t_q=0$, as in \cref{fig:constant-trajectory}(a). The quench time is $t_q=0.01/J$ for $N=2$ and $0.02/J$ for $N=3$ for a total measurement time of $t_0+t_q$. The dashed green curve shows data for $N=3$ with a longer quench time of $t_q=0.03/J$.
    \textbf{(b)} The trajectory of the controlled central pinning $\Delta\mu_6$ as a function of time, showing the optimised $t_0=0.1377/J$ ($0.0696/J$) and the quench time $t_q=0.01/J$ ($0.02/J$) for $N=2$ ($N=3$) as identified in panel (a). The evolution is stopped at time $t_0+t_q$.
    Data is for the BH model with $U=-10J$, and $M=11$.}
    \label{fig:tuning-t0}
\end{figure}

The protocol for preparing a quantum soliton described in Sec.~\ref{sec:preparation} is ideal since the chemical potential quench at the end of the protocol is assumed to be instantaneous. In this Appendix, we discuss a more realistic protocol that accounts for a finite-time quench: experimentally, there are limitations on the rate of variation of the transmon frequencies (that is, chemical potentials). Typically,  
$\mu_i$ can be approximately tuned at a rate of 1~GHz per 10--20~ns~\cite{Blais:2020wjs,Rol:2019bvv}. 
To account for this limitation, we ramp the pinning to zero in time $t_q$ to perform a finite-time quench.
Considering the magnitude of $\mu^\text{opt}$, we set $t_q=0.01/J$ ($0.02/J$) for $N=2$ ($N=3$). We then modify the ideal protocol by adapting the pinning hold time $t_0$ to incorporate the finite-time quench. Intuitively, we expect that the final infidelity with the projected stack would benefit from a pinning time shorter than $t_0^\text{opt}$, since some state evolution will still take place during the quench.

\cref{fig:tuning-t0}(a) shows the infidelity between the projected boson stack (for $N=2,3$) and the state at the final protocol time $t_0+t_q$ as a function of the pinning hold time detuning, $t_0-t_0^\text{opt}$, where $t_0^\text{opt}$ is as identified in \cref{fig:constant-trajectory}(b). The state is first evolved unitarily for $t\in[0,t_0]$ with the chemical potential fixed to the optimal pinning $\mu^\text{opt}$, followed by evolution under a time-dependent Hamiltonian characterized by the linear quench of the chemical potential $\Delta\mu_6(t)=\mu^\text{opt}[1-(t-t_0)/t_q]$ for $t\in[t_0,t_0+t_q]$ [cf. \cref{fig:tuning-t0}(b)]; during this time
the evolution is computed through Trotterization, see \cref{app:breathing}. 
The data at $t_0-t_0^\text{opt}=0$ shows that without tuning the hold time, the infidelity only slightly increases for $N=2$, but is about one order of magnitude larger for $N=3$. The infidelity is a nonmonotonic function of the pinning hold time and is minimized for $t_0<t_0^\mathrm{opt}$, as qualitatively argued above; 
notably, the fidelity loss due to the finite-time quench can be fully compensated for by reducing the hold time, see horizontal dashed lines. As the number of bosons increases, the optimal time $t_0^\text{opt}$ decreases and a longer quench time is required due to the larger value of $\mu^\text{opt}$. The dashed green line in \cref{fig:tuning-t0}(a) shows data for $N=3$ with a longer quench time of $t_q=0.03/J$, demonstrating that the effects of a slower finite-time quench can be compensated for by a further reduction in hold time. 

\bibliography{bib}

\end{document}